\documentclass[review]{elsarticle}

\usepackage{hyperref,graphicx}
\usepackage{color}

\journal{Carbon}

\begin{document}

\begin{frontmatter}

\title{The Significance of Bundling Effects on Carbon Nanotubes' Response to Hydrostatic Compression}


\author[mymainaddress]{Y.W.Sun\corref{mycorrespondingauthor}}
\cortext[mycorrespondingauthor]{Corresponding author}
\ead{yiwei.sun@qmul.ac.uk}

\author[mysecondaryaddress]{I.Hern\'{a}ndez}

\author[mysecondaryaddress]{J.Gonz\'{a}lez}

\author[mymainaddress]{K.Scott}

\author[mymainaddress]{K.J.Donovan}

\author[mymainaddress]{A.Sapelkin}

\author[mysecondaryaddress]{F.Rodr\'{\i}guez}

\author[mymainaddress]{D.J.Dunstan}

\address[mymainaddress]{School of Physics and Astronomy, Queen Mary University of London, London E1 4NS, UK}

\address[mysecondaryaddress]{Departamento CITIMAC, Universidad de Cantabria, 39005 Santander, Spain}

\begin{abstract}
The study of the G-mode pressure coefficients of carbon nanotubes, reflecting the stiff \textit{sp$^2$} bond pressure dependence, is essential to the understanding of their extraordinary mechanical properties as well as fundamental mechanics. However, it is hindered by the availability of carbon nanotubes samples only as bundles or isolated with surfactants. Octadecylamine functionalized carbon nanotubes are mostly of a single diameter and can be stably dispersed in 1, 2-dichloroethane and chloroform without surfactants. Here we perform high pressure Raman spectroscopy on these tubes and obtain their experimental G-mode pressure coefficients for individual tubes and bundles. The G$^{+}$ pressure coefficient for bundles is only about half of that for individual tubes in 1, 2-dichloroethane and is about two-thirds in chloroform. The G$^{-}$ pressure coefficient for bundles is about one-third of G$^{+}$ in 1, 2-dichloroethane and about the same in chloroform. These results for the first time provide unambiguous experimental evidence of the significant effect of bundling on carbon nanotubes' G-mode pressure coefficients, identifying it as one of the major reasons for the lack of consensus on what the values should be in the literature.
\end{abstract}


\end{frontmatter}

\section{Introduction}

Carbon nanotubes (CNTs) are known for their extraordinary mechanical properties. In particular, their exceptionally high Young's moduli in the terapascal range, resulting from the in-plane \textit{sp$^2$} bond between the carbon atoms, make CNTs the stiffest materials yet discovered \cite{Treacy1996}. To study and understand the huge resistance to compression, relating to the pressure dependence of the covalent \textit{sp$^2$} bond, Venkateswaran \textit{et al.} performed the first Raman spectroscopy experiment on CNTs under high pressure and got the pressure coefficients of their in-plane vibrational modes G-mode (GM) in 1999 \cite{Venkateswaran1999}. If not specified, GM refers to the dominant G$^{+}$ mode, vibrating along the tube axis, rather than the G$^{-}$ perpendicular to axis. The upshifts of CNTs GM frequencies with pressure are believed to be mostly induced by the \textit{sp$^2$} bond stiffening, as in the cases of graphite \cite{Hanfland1989} and graphene \cite{Proctor2009}. Therefore, it is reasonable to consider that the GM pressure coefficients should be similar for graphite, graphene and CNTs. Not considering some extreme environmental effects, graphene does have similar GM shift rates to graphite \cite{Hanfland1989,Proctor2009}, while CNTs have a wide range of values, reported by different research groups over the past 15 years \cite{Venkateswaran1999,Christofilos2007,Christofilos2005,Ghandour2011,Venkateswaran2001,Proctor2006,Thomsen1999,Lebedkin2006}. Ghandour \textit{et al.} attributed the disparity to the tube diameter and explained the diameter dependence of the GM pressure coefficients with a \emph{thick-wall tube model} \cite{Ghandour2013}. Such diameter dependence is expected, although the model itself is not perfect, requiring a lower graphene GM pressure coefficient than experimental value, and not taking recently reported chirality effects into account \cite{Sun2014}. Instead of measuring tubes of mixed diameters, Raman measurements now preferably involve tubes of a single diameter, which is mainly done in two ways: using resonance-enhanced Raman spectroscopy (RRS), which in principle requires a wavelength tunable laser or using CNTs samples of a single diameter. 

\begin{figure}
\includegraphics[width=\linewidth]{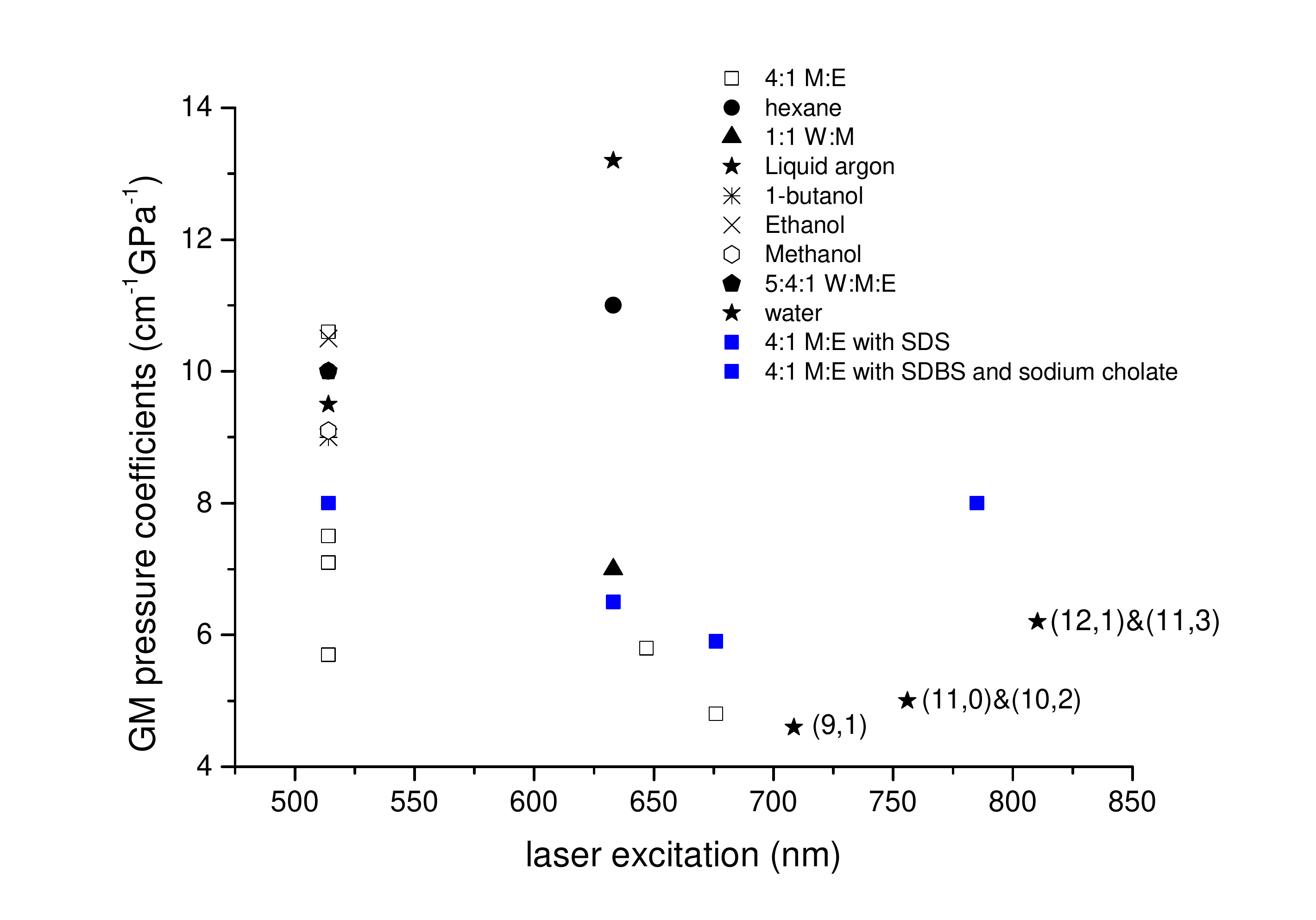}
\caption{CNTs GM pressure coefficients reported in previous literature are plotted against the laser excitation wavelengths, which they were obtained. Symbols identify the different PTMs. The blue squares are for individual tubes dispersed by surfactants. Three points are labelled with specific chiralities, to which they are assigned. M---Methanol; E---Ethanol; W---Water; SDS---Sodium dodecyl sulfate; SDBS---Sodium dodecylbenzene sulfonate.}
\label{GM}
\end{figure}

Most such experiments, as mentioned above, obtained different GM pressure coefficients from each other on mixed diameters samples at certain laser excitation wavelength in various pressure transmitting media (PTM), as shown in Figure~\ref{GM} \cite{Venkateswaran1999,Christofilos2007,Christofilos2005,Ghandour2011,Venkateswaran2001,Proctor2006,Thomsen1999,Lebedkin2006}. These are examples of close-ended tubes. The disparity comes both from the mixture of diameter and from environmental effects \cite{Ghandour2013}--- a specific environment (PTM, bundling and surfactants) and excitation wavelength picks out tubes of a certain diameter. A recent advance has been made in that experiments on CNTs in water enable the assignment of the observed GM pressure coefficients to tubes of particular chiralities (See Figure~\ref{GM}) \cite{Ghandour2013}. 

As well as intrinsic effects such as diameter and chirality, the pressure dependence of GM can be affected by exogenous effects, such as bundling, due to the van der Waals interaction between the tubes within a bundle. CNTs tend to form bundles \cite{Bandow1998}. Moreover, bundling effects on the GM pressure coefficient of the tube picked out by RRS may vary with parameters such as the diameters of the surrounding tubes, the bundling configuration (tangled, etc) and the degree of bundling, which is affected by the sample concentration but cannot be precisely controlled. 

On the other hand, surfactants stably disperse CNTs, which allows to exclude the bundling effects and their uncertainties, but possibly introduces surfactant effects (van der Waals interaction between the ambipolar surfactant molecules and CNTs). Researchers compared the GM shift rates of individual tubes to the ones of bundles \cite{Christofilos2007,Lebedkin2006}. It is worth noticing that though shifting the GM frequency, the contribution of the new ions in solution brought by surfactants does not affect much the CNTs pressure response (the GM pressure coefficients) \cite{Christofilos2007}. Again, in literature most research on tubes individualised by surfactants were done on samples of mixed diameters and reported varied values of GM pressure coefficients (see Figure~\ref{GM}).  

\begin{figure}
\includegraphics[width=0.5\linewidth]{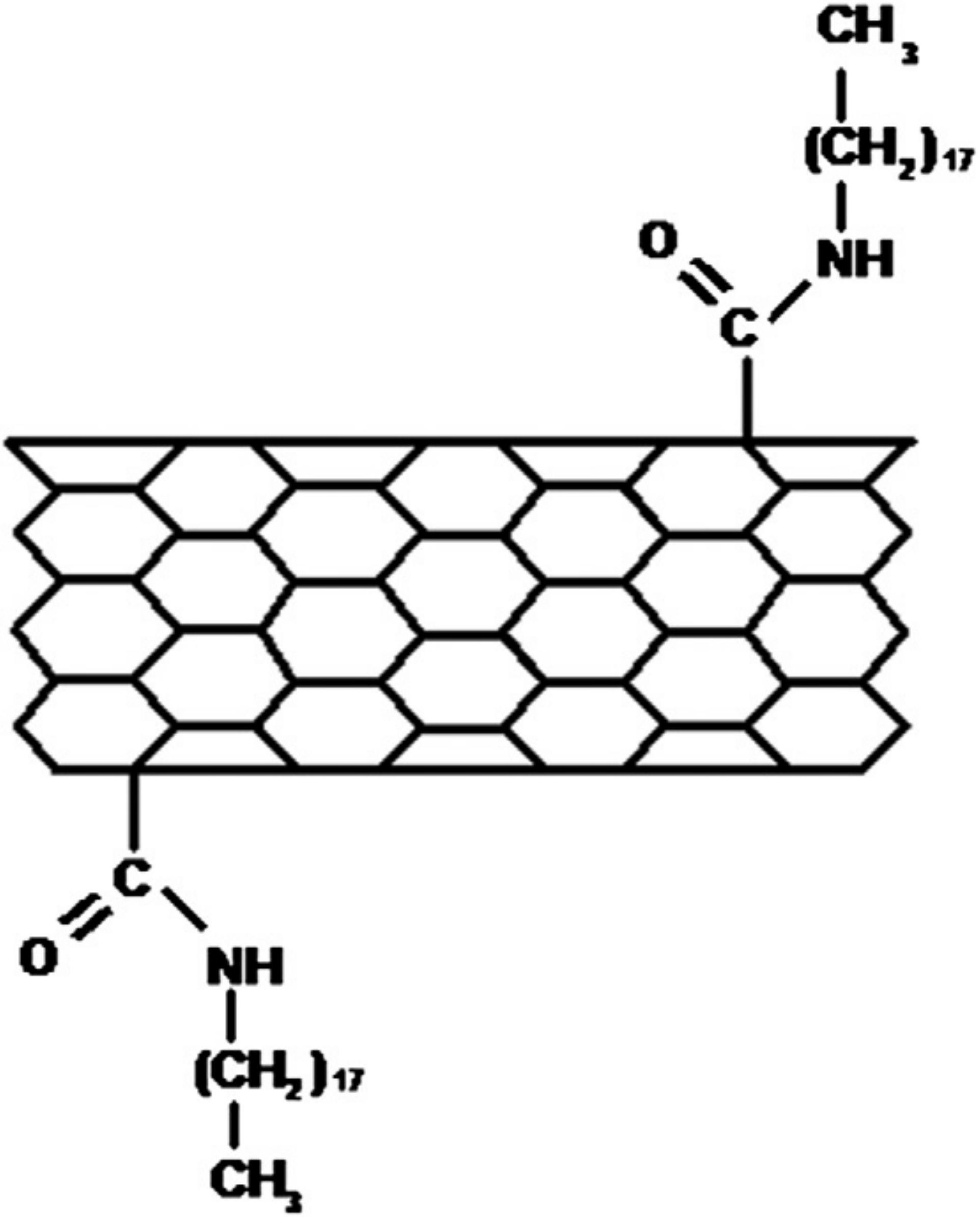}
\caption{A scheme of an ODA functionalized CNT.}
\label{ODA}
\end{figure}

Functionalized CNTs provide an alternative approach to study the GM pressure dependence, excluding both the diameter effects and the strong van der Waals interaction between tubes or between tube and surfactants, and that is the approach we use here, with octadecylamine (ODA) functionalized tubes. The functional groups are expected to keep the tubes apart in solution and deter bundling, by steric hindrance (see Figure~\ref{ODA}), while having themselves much less effect on the tubes than surfactant molecules since they are bonded to only one carbon in the order of a hundred. Venkateswaran \textit{et al.} first studied the pressure response of ODA functionalized tubes in 2001 \cite{Venkateswaran2001}. However, they used ODA tubes in solid (powder) form, which are still bundles (albeit small bundles). No further pressure experiments have been reported on ODA tubes since then. Here we carry out a complete and systematic study, clearly exposing the advantages and disadvantages of using such tubes.

The typical bundle diameter of the solid form of ODA functionalized tubes is 2--8 nm while the length is 0.5--1 $\mu$m, which Donovan \textit{et al.} considered to be a very low degree of bundling \cite{Donovan2012}. They can be stably dispersed in certain organic solvents, such as 1,2-dichloroethane (DCE) and chloroform, without the aid of surfactants. In fact, DCE disperses well even non-functionalized single-walled carbon nanotubes (SWCNTs) after sonication (this can be imaged by STM techniques \cite{Venema2000}), but the dispersion does not persist long enough for a series of Raman measurements under pressure \cite{Venema2000}. The steric hindrance caused by the functional group coverage of SWCNTs between 1.8 and 3.2 ODA chains per nanometre stabilizes the suspension \cite{Donovan2013}. It must be noted that as a result of acid treatment during the ODA functionalization, the caps at the end of tubes are removed \cite{Niyogi2002} and this raises the issue, whether the PTM can enter freely into the tubes. The density of states is largely disrupted \cite{Chen1998} and therefore no resonance condition applies. This has the advantage that the contribution of CNTs in a given sample to the Raman spectrum is independent of the laser excitation wavelength. 

It would be outside the scope of the work reported here to provide a clear answer to what the GM pressure coefficients should be, out of the various values reported in the literature. However, these experiments do present unambiguous experimental evidence, for the significant contribution that bundling alone makes to the pressure coefficients, by comparisons between tubes individualized without surfactants and bundles in DCE and chloroform.

\section{Experimental}

The ODA functionalized tubes were used as purchased from Carbon Solutions, who synthesized them by the arc discharge method, and functionalized them with ODA following a nitric acid treatment \cite{Chen1998}. The manufacturer specifies that the carbonaceous purity is over 90\%, in which SWCNTs loading is 65\%$\pm$15\%, determined by solution-phase near-IR spectroscopy \cite{Itkis2003}. When first synthesising ODA tubes by this route, Chen \textit{et al.} reported a single radial breathing mode (RBM) Raman peak at 170 cm$^{-1}$ in CS$_2$, at 1064 nm excitation wavelength \cite{Chen1998}, indicating that they are of single diameters 1.41 nm, according to the commonly used relation \cite{Maultzsch2005}.

\begin{equation}
d = \frac{215}{\omega_{RBM}-18}
\label{eq1}
\end{equation}

We prepared four samples of ODA functionalized CNTs --- bundled SWCNTs (b-SWCNTs) and individual SWCNTs (i-SWCNTs) in DCE, b-SWCNTs and i-SWCNTs in chloroform, following the recipe (sonication time, power, etc.), which Donovan \textit{et al.} used in their study \cite{Donovan2012,Donovan2013}. The dispersion was tested by the dichroism \cite{Donovan2012} and viscosity of the solution \cite{Donovan2013}, determined by polarizability, thus sensitive to the bundled or individualized status. The concentrations of the samples were 1$\times$10$^{-4}$ wt\% for b-SWCNTs in DCE, 1$\times$10$^{-6}$ wt\% for i-SWCNTs in DCE, 1.5$\times$10$^{-4}$ wt\% for b-SWCNTs in chloroform and 1$\times$10$^{-6}$ wt\% for i-SWCNTs in chloroform.  

Room temperature non-polarized Raman spectra of the samples were obtained in the backscattering geometry with a Horiba T64000 Raman system with a confocal microscope that had a resolution of 0.6 cm$^{-1}$, a single 1800 grooves/mm grating and a 100-$\mu$m slit, and was equipped with a liquid N$_2$-cooled CCD detector (Jobin-Yvon Symphony). Suitable edge filters for the 488 nm, 514 nm and 647 nm lines of a Coherent Innova Spectrum 70C Ar$^{+}$-Kr$^{+}$ laser could be used with the system. We kept the laser power on the sample below 5 mW to avoid significant laser-heating effects on the probed material and the concomitant softening of the Raman peaks. 

For the high pressure experiments, we used a membrane diamond anvil cell with anvils of 500 $\mu$m culet size and very low fluorescence (Type IIa). The ruby luminescence R1 line was used for pressure calibration \cite{Mao1986}. For the Raman spectroscopy, we used a 20$\times$ objective on the b-SWCNTs, i-SWCNTs in DCE and i-SWCNTs in chloroform in the pressure cell and a 40$\times$ objective on b-SWCNTs in chloroform. We set the pinhole size in confocal configuration at 200 $\mu$m. These settings were found to give the best quality spectra. The four samples were separately loaded into the cells in four separate experiments. After loading the samples into cells with a small pressure applied to prevent the solvent from evaporating, initial RBM and GM spectra of all the samples under 488, 514 and 647 nm excitation were obtained. Then the Raman was measured at higher pressures under 488 nm excitation for b-SWCNTs in DCE and b-SWCNTs in chloroform, and under 514 nm excitation for i-SWCNTs in DCE and i-SWCNTs in chloroform. In all cases, the signal-to-noise ratio of the GM spectra decreased with pressure and therefore this study is in a low pressure range, well below 10 GPa, which is the reported experimental collapse pressure of CNTs of diameters similar to those used here \cite{Caillier2008}. 

The hydrostaticity in this high pressure experiments has been studied. DCE solidifies at 0.6 GPa \cite{Sabharwal2007} and chloroform between 0.60--0.79 GPa \cite{Dziubek2008}. We obtain very similar ruby R1 wavelength from two different ruby chips in the cell: 694.66 and 694.67 nm at 1.20/1.24 GPa, 694.86 and 694.87 at 1.74/1.76 GPa for b-SWCNTs in chloroform, after observing the solvent solidification. The corresponding errors in pressure are 3.3\% and 1.1\%, showing acceptable hydrostaticity in this high-pressure study.

\section{Results and Analysis}

\begin{figure}
\includegraphics[width=0.9\linewidth]{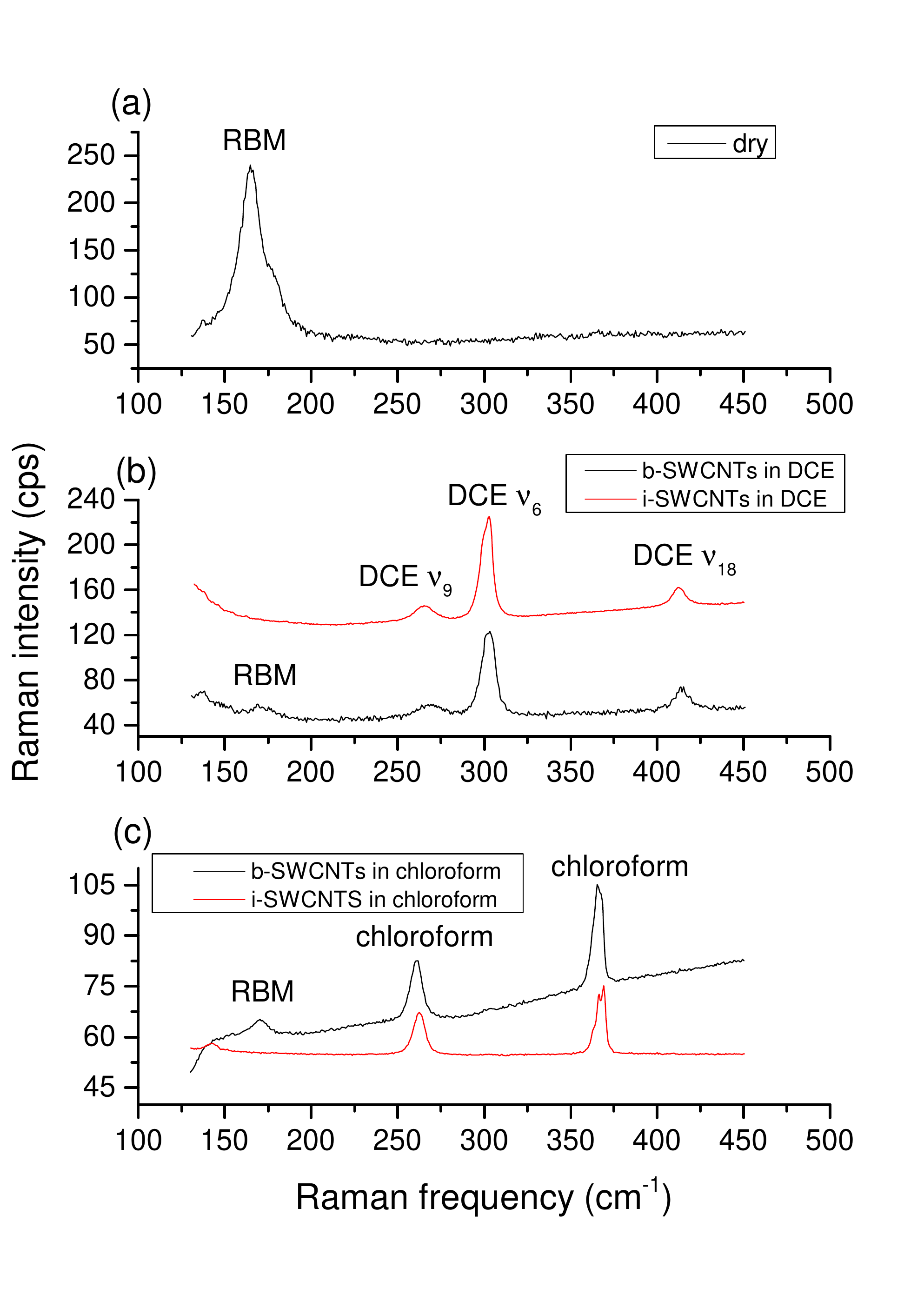}
\caption{The RBM spectra of ODA functionalized tubes are shown for (a) dry samples on a glass slide, (b) b-SWCNTs (black) and i-SWCNTs (red) in DCE and (c) b-SWCNTs (black) and i-SWCNTs (red) in chloroform. In (b) and (c) the spectra are vertically shifted for clarity. The Raman peaks from the solvent are labelled. Laser excitation wavelengths are 488 nm for b-SWCNTs in DCE and b-SWCNTs in chloroform, and 514 nm for i-SWCNTs in DCE and i-SWCNTs in chloroform. Raman shifts do not vary with the excitation wavelength.}
\label{RBM}
\end{figure}

Figure~\ref{RBM} shows the raw RBM spectra of the dry sample on a glass slide and four prepared samples in cells. The spectra are vertically shifted for clarity to compare the spectra of i-SWCNTs to b-SWCNTs.

Following the literature, we assign the peaks at 268, 302 and 414 cm$^{-1}$ to DCE \cite{Sabharwal2007} and the peaks at 251 and 368 cm$^{-1}$ to chloroform \cite{Hubel2006}. For the CNTs, the fitted RBM frequencies (Lorentzian fit) of the dry sample are at 164.9 and 179.5 cm$^{-1}$, and correspond to tubes of diameters 1.46 and 1.33 nm, according to Eq.~\ref{eq1}. The ratio of the peaks' integrated area is 9.85:1, former to latter. The small RBM peak cannot be detected for samples loaded into diamond-anvil cells as the absorption by diamonds weakens the signal. For i-SWCNTs which are at an order of magnitude lower concentration than b-SWCNTs, even the main peak is no longer detectable. 

\begin{figure}
\includegraphics[width=\linewidth]{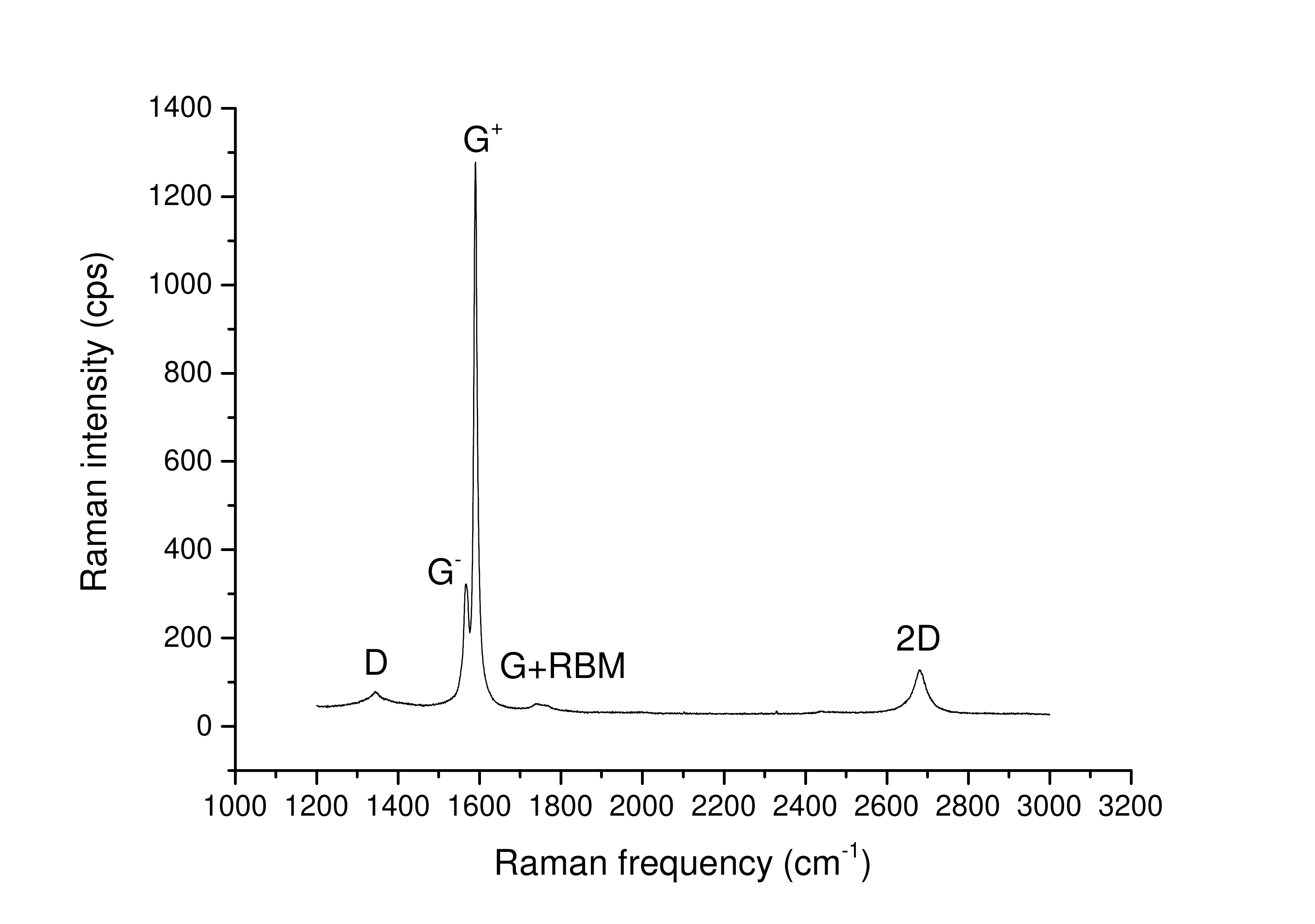}
\caption{The D, G and 2D spectrum of ODA functionalized tubes are shown for dry samples on a glass slide. The laser excitation wavelength is 514 nm.}
\label{D}
\end{figure}

Figure~\ref{D} shows the raw D, G and 2D-band spectra of the dry sample on a glass slide. A clear single G$^{-}$ peak at 1565.9 cm$^{-1}$ can be observed. The defectiveness of the tubes can be judged by the peak intensity ratio of the G to D-band features $I_G/I_D$=46.33. This may be compared with the values given by Brown \textit{et al.} for non-functionalized SWCNTs of a diameter distribution vary from about 2 to 30 \cite{Brown2001}. The low implied defectiveness is not unexpected given that the coverage of the functional groups is between 1.8 and 3.2 ODA chains per nanometre (approximately per 150 carbon atoms). Thus we suppose that the GM and RBM of these tubes are representative of the unperturbed (non-functionalised) tubes. 

Before presenting the data, it may be noted that the signal to noise ratio of the Raman spectra presented here is low, for two reasons. Firstly, the Raman experiments are performed under non-resonance conditions as the density of states of these tubes is largely disrupted and thus the peak intensities are up to 6 orders of magnitude lower than those under resonance conditions. Secondly, extremely low samples concentration are used (see 'Experimental'), which is necessary for the stability of the dispersions.

\begin{figure}
\includegraphics[width=\linewidth]{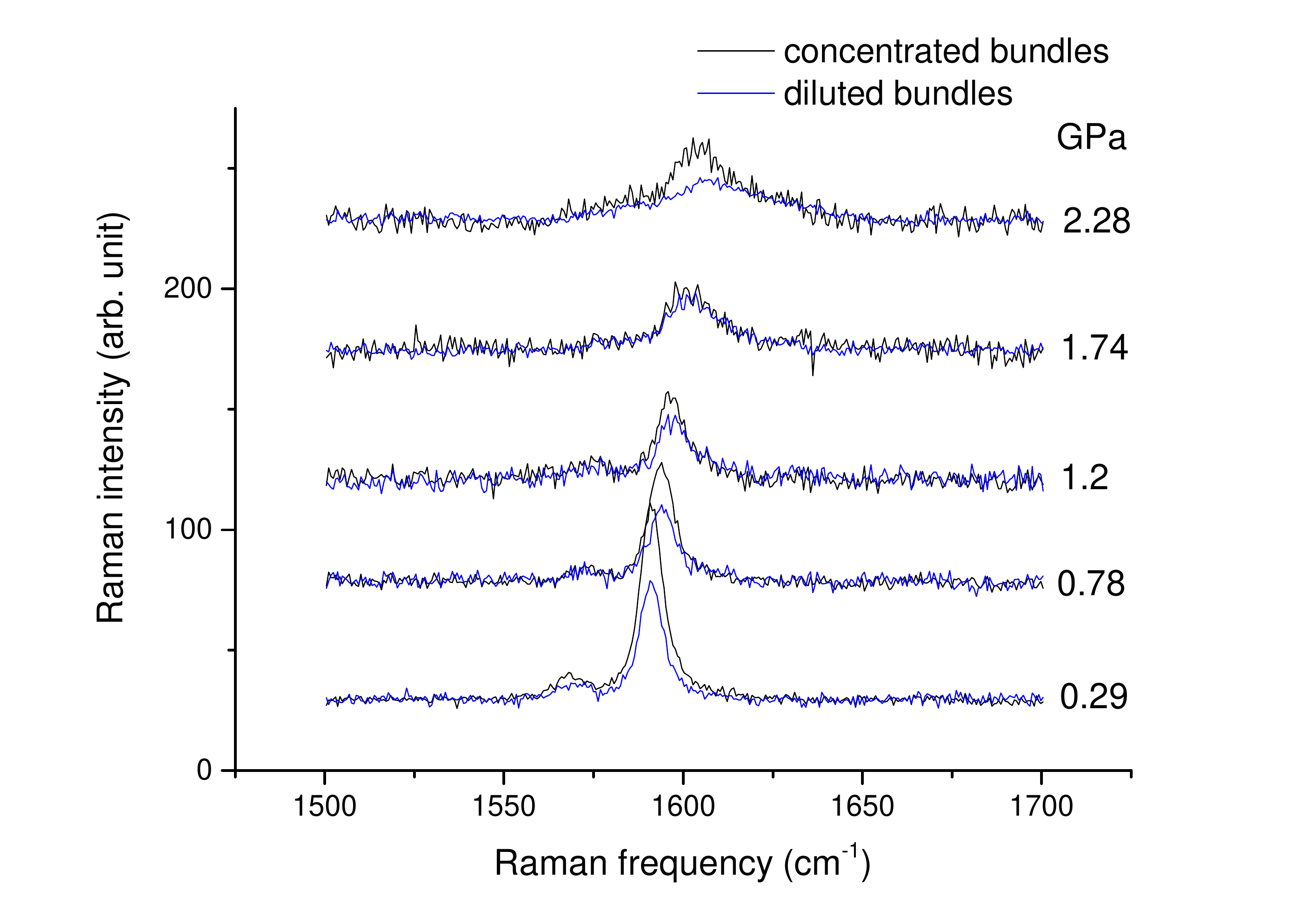}
\caption{GM spectra of b-SWCNTs in chloroform are collected from the dark area (black) and the transparent area (blue). The spectra are vertically shifted, proportional to pressure. The pressures, under which the spectra are obtained, are labelled. The laser excitation wavelength is 488 nm.}
\label{l_s}
\end{figure}

We need to consider the effect of concentration on CNTs GM pressure coefficients, as i-SWCNTs and b-SWCNTs, which we are going to compare, are of different concentration. Figure~\ref{l_s} presents the GM spectra obtained in high pressure measurements on b-SWCNTs in chloroform, from two different spots --- one in a dark area, which is richer in bundles, making it observable under microscope, and the other in a transparent area, which is less concentrated. We label the GM spectra as concentrated bundles and diluted bundles. The baselines are subtracted, and then the spectra are vertically shifted, proportional to pressure. Importantly, figure~\ref{l_s} shows that the GM frequencies are nearly unaffected by the sample concentration and therefore it is reasonable to consider that the GM pressure coefficients are independent of the sample concentration in the low pressure range in this study.

\begin{figure}
\includegraphics[width=0.9\linewidth]{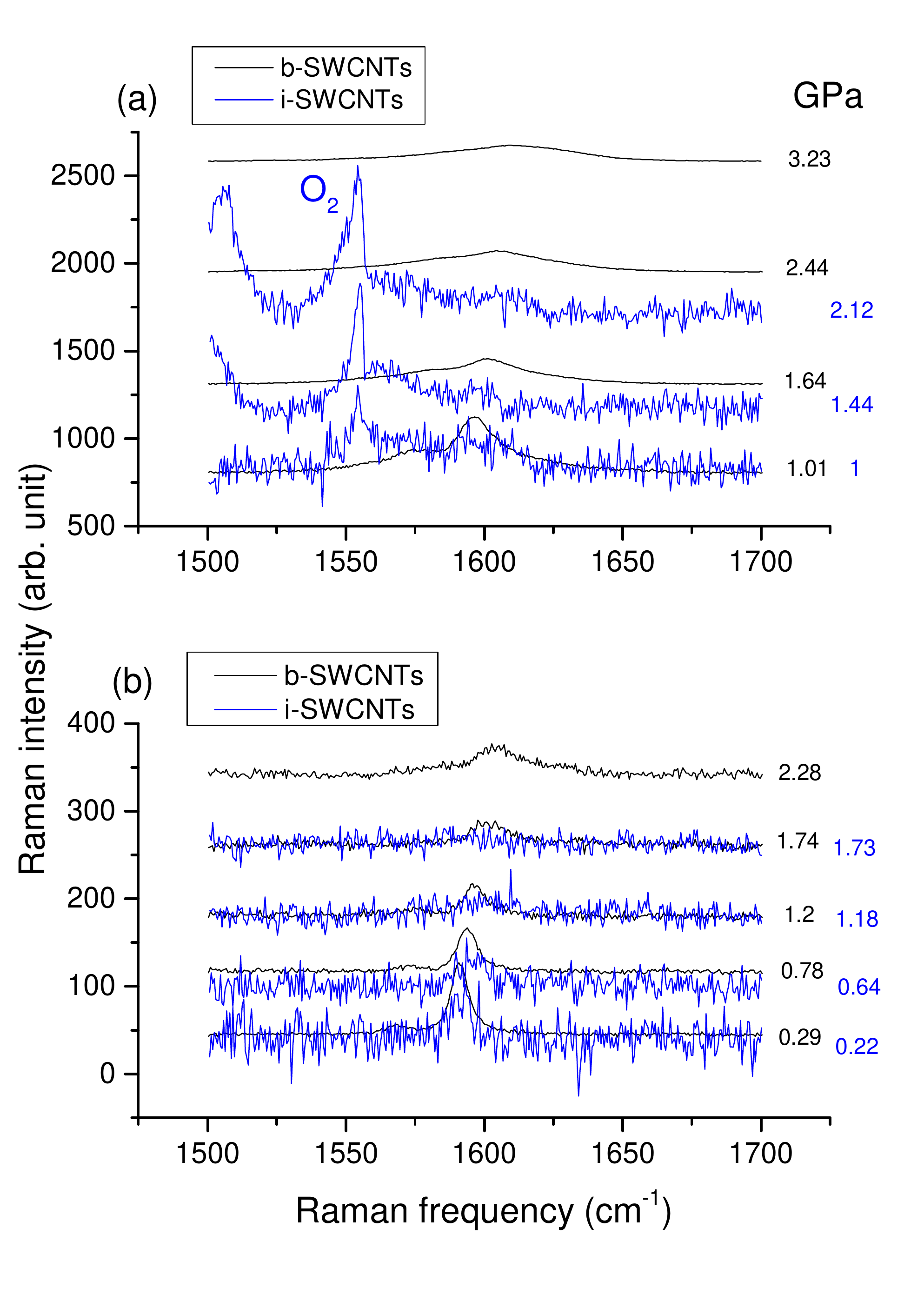}
\caption{The GM spectra of i-SWCNTs (blue) and concentrated b-SWCNTs (black) are shown in (a) DCE and (b) chloroform. For i-SWCNTs the Raman intensities are multiplied by 200. The spectra are vertically shifted, proportional to pressure. The pressures, under which the spectra are obtained, are labelled in the colours corresponding b-SWCNTs (black) or i-SWCNTs (blue). The oxygen vibrational Raman peaks are labelled. The laser excitation wavelength is 488 nm for b-SWCNTs in DCE and b-SWCNTs in chloroform, and 514 nm for i-SWCNTs in DCE and i-SWCNTs in chloroform. Raman shifts do not vary with the excitation wavelength.}
\label{i_b}
\end{figure}

Figure~\ref{i_b} exhibits the GM spectra of b-SWCNTs and i-SWCNTs at similar pressure points in both DCE and chloroform. For b-SWCNTs in chloroform, the spectra are those of concentrated bundles in Figure~\ref{l_s}. For i-SWCNTs in DCE and chloroform, the Raman intensities are multiplied by a factor 200, in order to get clear comparisons to the b-SWCNTs. As a result, the i-SWCNTs spectra show an increased level of noise compared to b-SWCNTs spectra. The baselines are subtracted, and then the spectra are shifted vertically, proportional to pressure. The narrow peak at 1554.4 cm$^{-1}$ in the GM spectrum of i-SWCNTs in DCE is assigned to an oxygen vibrational Raman peak from the air between the microscope and the cell \cite{Fletcher1974}. It is on top of a wide peak, which might be from carbonaceous impurities in the samples. The GM peak is right next to the wide peak and the signal to noise ratio is low. 

\begin{figure}
\includegraphics[width=\linewidth]{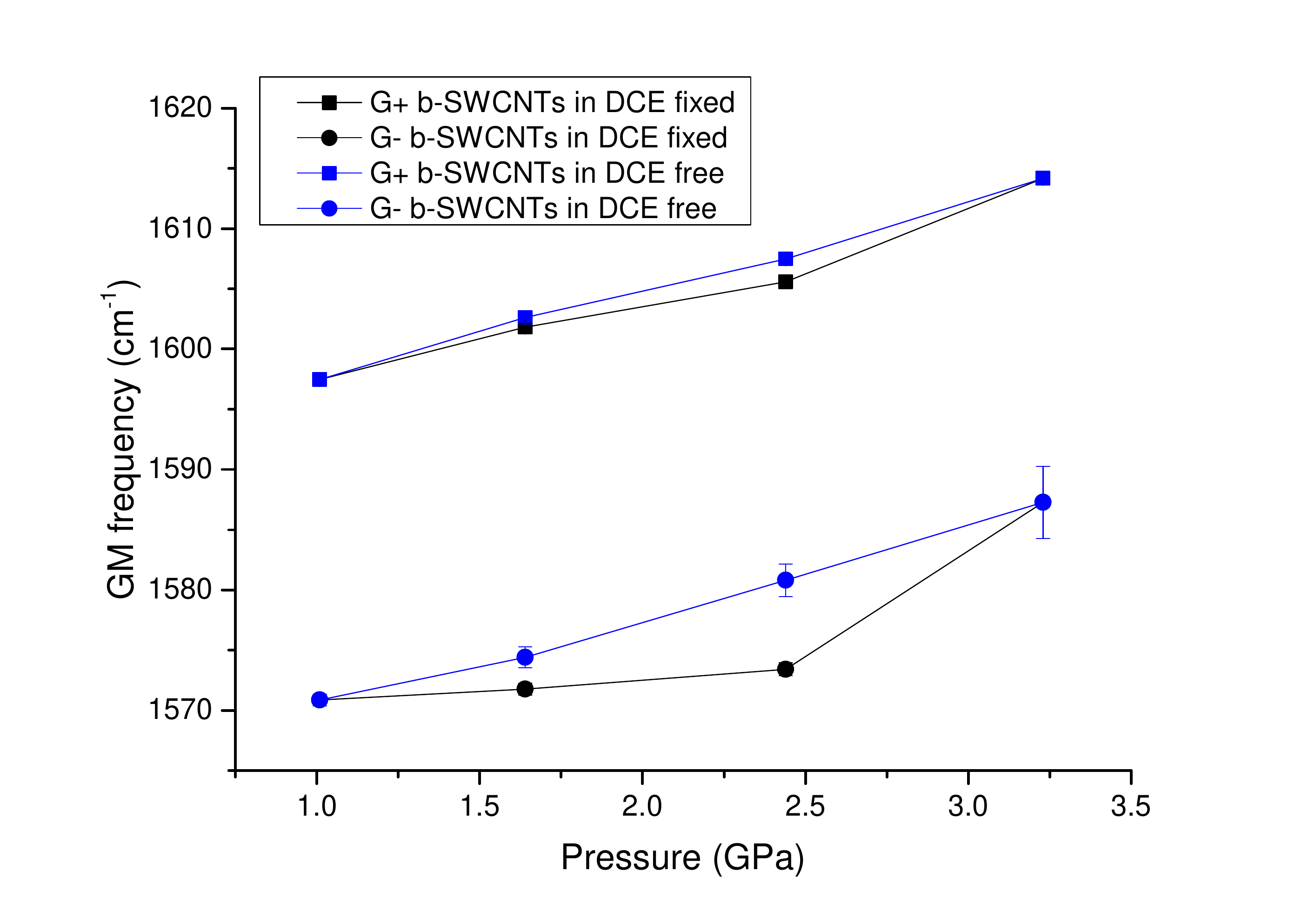}
\caption{The GM frequencies of b-SWCNTs in DCE  are plotted against pressure. The frequencies are obtained by fixing the integrated area ratio of G$^{+}$ to G$^{-}$ at the free fitting value at 1.01 GPa (black), and by releasing all the fitting parameters of the two Lorentzians (blue). The squares are for G$^{+}$ and the circles are for G$^{-}$. Where they exceed the size of the data-points, error bars are shown.}
\label{fitting}
\end{figure}

\begin{figure}
\includegraphics[width=\linewidth]{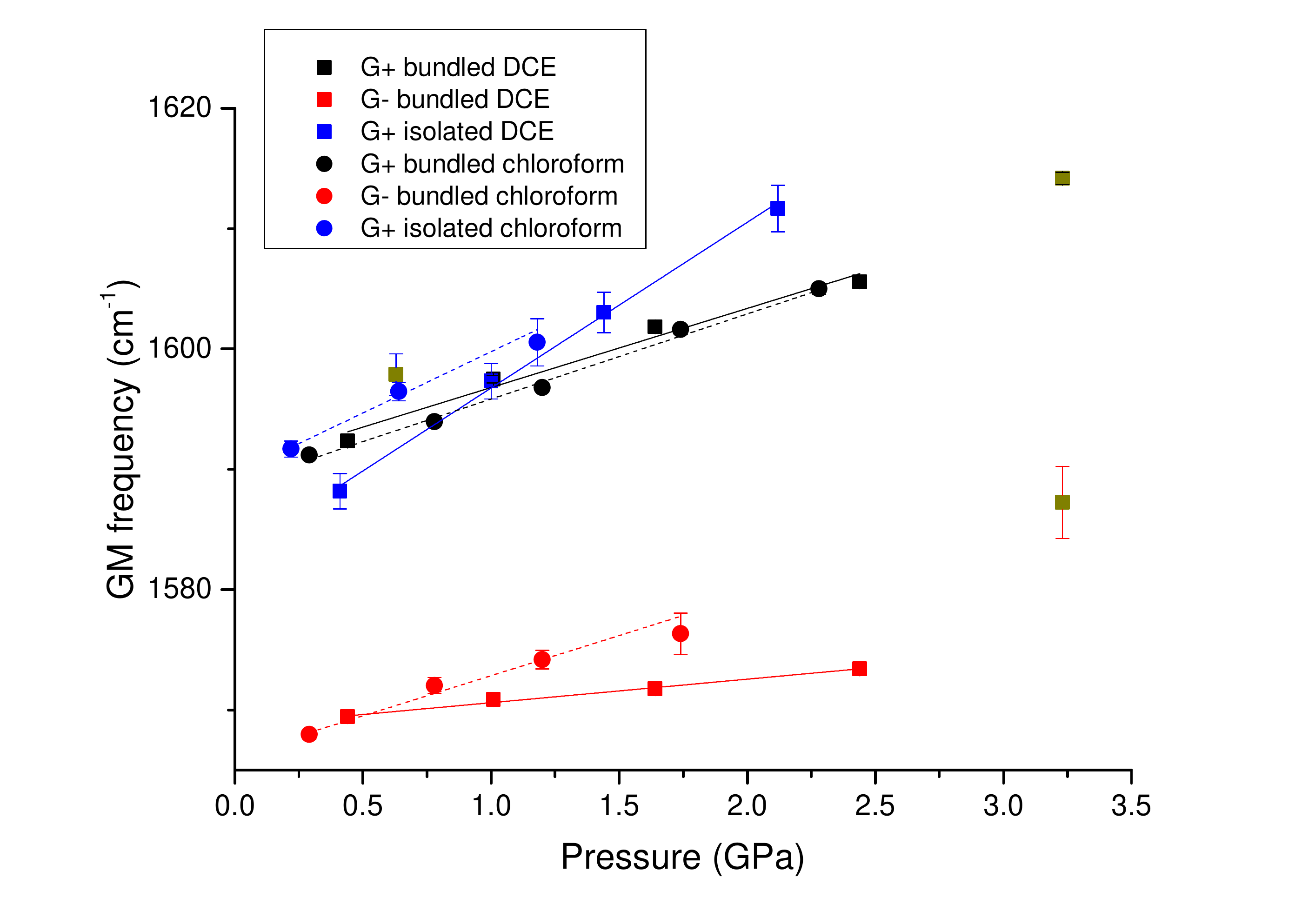}
\caption{The GM frequencies of all the samples are plotted against pressure. The colour black is for G$^{+}$ frequencies of b-SWCNTs, red is for G$^{-}$ of b-SWCNTs and blue is for G$^{+}$ of i-SWCNTs. The squares are for samples in DCE and the circles are for chloroform. Error bars are shown, where they exceed the size of the points. The linear fits are presented as solid lines for DCE and dashed lines for chloroform. The fits exclude the points for b-SWCNTs in DCE at 3.23 GPa and i-SWCNTs in DCE at 0.63 GPa, which are shown as open symbols.}
\label{results}
\end{figure}

We fit the GM spectra of i-SWCNTs in DCE and chloroform in Figure~\ref{i_b} each with a single Lorentzian and the GM spectra of b-SWCNTs in DCE and chloroform in Figure~\ref{i_b} each with two Lorentzians. In the latter case these correspond to the G$^{+}$ and G$^{-}$ peaks, which are initially well separated but cannot be told apart with increased pressure. We fix the integrated area ratio of G$^{+}$ to G$^{-}$ at the value obtained by free fitting at the first pressure point during the whole fitting, to avoid unphysical fitting results such as a larger G$^{-}$ than G$^{+}$ peak, that may be obtained when releasing all the fitting parameters of the two Lorentzians. Figure~\ref{fitting} shows how these two fitting procedures lead to different GM frequencies of b-SWCNTs in DCE. The difference is mainly at the uncertain frequencies of the weak G$^{-}$ peak. We plot GM frequencies, obtained by fixing the integrated area ratio, of all the samples against pressure with error bars in Figure~\ref{results}. Linear least square fits are shown, excluding the points of b-SWCNTs in DCE at 3.23 GPa, for the similar pressure range to that of i-SWCNTs in the same solvent and the point of i-SWCNTs in DCE at 0.63 GPa as an abnormal point, which is exactly at the DCE solidification point. The excluded data points are labelled green.

We present the GM pressure coefficients in Table~\ref{table} from the linear fit in Figure~\ref{results}. The errors are from the linear fit, the Lorentzian fit for the peak position and the system resolution. 

Figure~\ref{results} and Table~\ref{table} present the key result that the G$^{+}$ pressure coefficient for bundles is only about half of that for individual CNTs in DCE and is about two-thirds in chloroform. For bundles, the G$^{-}$ pressure coefficient is about one-third of G$^{+}$ in DCE and about the same in chloroform.

\begin{table}
\centering
\caption{GM pressure coefficients for all measured samples} \label{table}
\begin{tabular}{l*{1}{c}r}
\hline\hline
GM pressure coefficients (cm$^{-1}$GPa$^{-1}$)  &G$^{+}$          &G$^{-}$ 		 \\
\hline
b-SWCNTs in DCE                                 &6.6$\pm$0.7 &2.0$\pm$0.1 \\
i-SWCNTs in DCE									&13.8$\pm$0.6 \\
b-SWCNTs in chloroform							&7.1$\pm$0.3 &6.7$\pm$0.7 \\
i-SWCNTs in chloroform							&10.2$\pm$1.3 \\
\hline\hline
\end{tabular}
\end{table}

\section{Discussion}
At the moment we do not fully understand these results. There are four key issues. First, the pressure dependence of GM is commonly considered as determined by the shortening of \textit{sp$^2$} bond, which should have little to do with the environment, in contract with the apparent bundling and solvent effects reported here. This is quite unlike the pressure dependence of RBM, which is due to the decrease of the distance between tube shell and the absorbed fluid layer and therefore unsurprisingly sensitive to the environment \cite{Longhurst2007}. Second, according to the \emph{thick-wall tube model} \cite{Ghandour2013}, the G$^{+}$ and G$^{-}$ pressure coefficients of 1.46 nm tubes should be 6.0 and 8.0 cm$^{-1}$GPa$^{-1}$ respectively. The model is based on individual tubes but the G$^{+}$ values of i-SWCNTs are much higher than the predicted ones. The G$^{+}$ values of b-SWCNTs agree well, as the previous work on bundles \cite{Ghandour2013}. Third, the tangential stress is always larger than the axial stress for a tube under hydrostatic pressure. The pressure coefficient of the vibrational mode along tube circumference (G$^{-}$) should therefore be always larger than the one along tube axis (G$^{+}$). This is again against our observations in DCE. And fourth, considering that the end of the tubes has been removed, the solvents might be expected to enter inside the tubes. If the internal pressure (pressure of the solvent inside the tube) is at a value between 0 and the external pressure, the pressure coefficients for both G$^{+}$ and G$^{-}$ should lie between the graphene value and the \emph{thick-wall tube model} predictions. Results of i-SWCNTs in Table~\ref{table} are out of this range. The normal way to judge whether tubes are solvent-filled by the shift of RBM frequency \cite{Cambre2010} is not possible in this case, because no close-ended ODA functionalized CNTs are available for comparison (the caps are removed during the ODA functionalisation).

ODA functionalized CNTs, as samples to study the CNTs GM pressure coefficients, have the following advantages. First, given that they are mostly of a single diameter and their density of states is largely disrupted, the contributions to the RBM and GM signals from tubes of different diameters may be taken proportional to their contents in samples, regardless of the laser excitation. In Figure~\ref{RBM}, we obtained the RBM integrated area ratio of 1.46 nm to 1.33 nm tubes at 9.85 to 1. The G$^{+}$ signal is contributed by 1.46 and 1.33 nm tubes, and in the absence of resonance, also with a ratio of 9.85 to 1. It is reasonable to attribute the GM pressure coefficients in Table~\ref{table} to 1.46 nm tubes only. Second, the ODA side chains offer the steric hindrance and therefore provide us with SWCNTs samples stably dispersed without the aid of surfactants. This is the main reason we use ODA CNTs in this study.

There are related disadvantages, namely the limited choices of PTM consistence with dispersion and the potential side chain effect on GM pressure coefficients. DCE and chloroform are effective in dispersing CNTs samples but are not considered as good PTMs because of their low solidification pressures. 

In order to exclude the inter-tube or tube-surfactant van der Waals interaction, we introduce the side chains. The ODA coverage between 1.8 and 3.2 chains per 150 carbon atoms may be high from the point of view of chemistry, but it is too low to have an effect on the in-plane vibrational frequencies at ambient pressure, and there is no reason to suppose it should have any more effect at high pressure. The upshift of GM frequency with pressure is induced by the increasing overlap of electrons of carbon atoms. In the case of bundles or surfactants, each carbon atom is under the influence and the behaviour of its electrons are affected, as shown in this work, while in the case of ODA tubes, electrons of most carbon atoms are not affected by the \textit{sp$^3$} defects (1.8--3.2 \textit{sp$^3$}s in 150 \textit{sp$^2$}s). Thus its effect on the pressure coefficient should be small, certainly not comparable to the effects of bundles or surfactants. Therefore, the experimental data of ODA functionalized tubes presented here can be meaningfully compared to the theories of the in-plane bond response to pressure in pristine SWCNTs.

\section{Conclusion}
In this work, we present the experimental demonstration of the significant and unexpected bundling effects on CNTs GM pressure coefficients of a specific chirality. The G$^{+}$ pressure coefficient for bundles is only about half of that for individual CNTs in DCE and is about two-thirds in chloroform. For bundles, the G$^{-}$ pressure coefficient is about one-third of G$^{+}$ in DCE and about the same in chloroform. Such comparison for the first time excludes the effect of surfactants, achieved by using ODA functionalized tubes. The origin of the bundling and solvent effects on GM pressure coefficients is unclear at the moment and the values of the pressure coefficients in this work are beyond the framework of the current understanding, especially the \emph{thick-wall tube model}. Despite posing unresolved questions, this work clarifies that the bundling effect, is one of the major reasons for the current lack of consensus on the value of GM pressure coefficients.

\section*{Acknowledgement}

The authors thank Prof. Alfonso San Miguel from University Lyon 1 for useful discussion on the way to present the experimental results. YWS thanks the Chinese Scholarship Council (CSC) for financial support and IH thanks the EU-P7 Marie-Curie Programme (CIG 303535).

\section*{References}

\bibliography{mybibfile}

\end{document}